\documentclass[10pt]{article}
\usepackage{longtable}
\usepackage{tabu}

\usepackage[hyphens]{url}
\usepackage{scrextend}
\usepackage[hidelinks]{hyperref}
\usepackage{amssymb}
\usepackage{tabularx}
\usepackage{booktabs}  
\usepackage[margin=1in]{geometry}  
\usepackage{graphicx}
\usepackage[ruled,linesnumbered]{algorithm2e}
\usepackage[numbers, sort]{natbib}
\usepackage[singlelinecheck=false]{caption}
\DeclareCaptionLabelFormat{blank}{}

\usepackage{booktabs} 
\usepackage{authblk}
\usepackage{adjustbox}
\usepackage{color, colortbl}
\usepackage{subcaption} 
\usepackage{amsthm}
\usepackage{amsmath}

\usepackage{pdflscape}
\usepackage{enumitem}
\usepackage{lineno}
\usepackage{float}
\usepackage[justification=centering]{caption}
\usepackage{makecell}
\usepackage[title]{appendix}

\definecolor{Gray}{gray}{0.9}
\definecolor{White}{rgb}{1,1,1}
\definecolor{LightCyan}{rgb}{0.88,1,1}
\definecolor{orange}{rgb}{1,0.5,0}

\usepackage[space]{grffile}

\makeatletter
\g@addto@macro{\thm@space@setup}{\thm@headpunct{:}}
\makeatother

\begin{document}
\title{Gender Gap Analysis in News and Talk Online Radio Broadcast}

\author{Gregory Koushnir\thanks{koushgre@post.bgu.ac.il}}
\author{Michael Fire\thanks{mickyfi@post.bgu.ac.il}}
\author{Dima Kagan\thanks{kagandi@post.bgu.ac.il}}
\author{Galit Fuhrmann Alpert\thanks{fuhrmann@bgu.ac.il}}

\affil[1]{Faculty of Computer and Information Science, Ben-Gurion University of the Negev, Israel}
    \maketitle

\begin{abstract}
Radio broadcasting remains a dominant medium of communication, reaching 82\% of Americans, ages 12 and older, weekly. Given its massive media impact, gender representation on radio news and talk stations plays a substantial role in shaping social and cultural perceptions. In this study we explored patterns of gender representation in radio broadcasts. Specifically, we focused on gender-based differences in total speaker allotted duration, air-time allocation at different times of day, and biases in gender participation across different broadcast topics. 

Our data set comprises filtered recordings from 74 US news and talk radio stations, collected over a period of 24 hours, yielding more than 1,400 hours of content. 
We applied on the data the VANPY analysis framework, an in-house comprehensive methodology that we developed to combine multi-channel radio recording with advanced AI processing for speaker diarization, gender classification, and topic analysis. 

Our results revealed consistant gender differences in the allocated broadcast time, with male speakers dominating  77\% (SD = 6.8\%) of the total speaking time.
This gender gap was even more prominent during peak broadcast times, with female representation remained approximately 7.5-9.5 minutes per hour during commute times, compared to approximately 30.6-32.9 minutes for male speakers. Gender differences were further demonstrated by topic analysis, using NLP-based classification of transcribed content. Female speaker representation was lowest in "Talk Show Segments", where women accounted for only 10.8\% of speaking time. Even within "Entertainment News" broadcasts, where female representation was highest compared to all other topics, they were only allocated as 35.8\% of speaking time. 

In addition to the important insights revealed in this study on gender dynamics in broadcast media,  
our in-house developed VANPY framework is publicly available  and provides a general and systematic applicable approach for the analysis of extensive audio data. VANPY could be used to explore speaker characteristics not only in radio broadcasts, but also in multiple other fields such as human-computer interaction, corporate communications, and security applications.

\end{abstract}
	\providecommand{\keywords}[1]{\textbf{Keywords:} #1}

\section{Introduction}
\label{sec:int}

Radio broadcasting has been a significant medium of communication for more than a century, serving as an approachable platform for the immediate consumption of news, music, and various forms of entertainment. 
Radio broadcasts provide a wealth of rich and dynamic content, reflecting a wide range of socio-cultural dynamics, news, entertainment, and advertising~\cite{jauert1997local, Sujoko31122023}. As such, they play a critical role in the dissemination of information and shaping public opinion~\cite{barker2000political}. 
Recent data confirms that radio continues to be dominant in audio consumption. In the United States, radio accounted for 64\% of daily ad-supported audio time in Q2 2025, compared with 19\% for podcasts, and 14\% for streaming audio services, and 3\% for satellite radio~\cite{Nielsen_2025_Q2}.
In the United States alone, according to 2023 reporting, radio content reaches approximately 82\% of adults weekly, having access to a variety of more than 15,400 stations~\cite{radio_topic}. 
A study from 2022 by~\citet{laor2022radio} reported that on-demand radio has successfully adapted to modern consumption patterns, with 76.5\% of questionnaire responders listen to online radio at least once weekly and about 60\% consume radio content on demand, accessing archived programs via website or mobile application. 

With tens of thousands of available radio stations, they naturally vary greatly, yet their formats can roughly be categorized into two main types - Speech and Music~\cite{shane1995modern} that are offered either separately or as a mix of both~\cite{warren2004radio, radio_2019_nielsen}. 
Speech content spans news, call-ins, topical discussions, sports, late-night shows, religious programs, or specialized information~\cite{warren2004radio, bonini2014radio}; while music content, typically, covers a wide variety of genres~\cite{nielsen_rating, radioformats_glenn}. Regardless of the format, commercial radio stations often also incorporate advertisements, either as clean voice inserts or accompanied by background music~\cite{ala2005saleable,sweeting2006coordination,generali2011happens}.
This rich diversity of content, combined with radio's widespread reach and recent digitization, makes radio broadcasts a valuable source for research and analysis across multiple dimensions. Specifically, a comprehensive analysis of radio broadcasts has the potential to highlight valuable insights into social issues such as gender representation, emotional expression, linguistic characteristics, and even non-verbal elements such as sound effects and applause~\cite{doi:10.1016/j.brq.2014.06.001, hurtz2004effects, Grekow2018, baksh2022said}.

In recent years, the analysis of radio data has drawn increased attention, driven by technological progress in audio processing and machine learning techniques~\cite{AudioandTextSentimentAnalysisofRadioBroadcasts, knowledge2030020, álvarez2024radiaradioadvertisement}. Machine Learning (ML) and Deep Learning (DL) methods are now being widely used to extract insights from audio data, including speech characteristics, speaker identification, and emotion recognition~\cite{purwins2019deep, JAHANGIR2021114591, SHAHFAHAD2021102951}. These techniques extract multiple features about a speaker, based solely on audio input, that can be used to describe the style and personality of speakers, as well as their psychological and physical states. Examples include gender classification~\cite{shafran2003voice, muller2006automatic}, age estimation~\cite{muller2006automatic}, accent detection~\cite{5670700}, emotional state classification~\cite{app12010327}, vocal pathology detection~\cite{kumar2016efficient} and speech style analysis~\cite{reid2022development, sheikh2022introducing}.

Previous voice analysis studies have documented persistent gender disparities in media representation~\cite{Kagan2020}, showing that female voices may be underrepresented both in terms of air-time and topic diversity~\cite{crider2020tomatoes, Williamson03072021, pelloin2024automatic}.
Building on this previous work, our study specifically examines how such disparities manifest in broadcast radio, using large-scale automatic analysis to explore gender differences in speaker representation and participation across news and talk formats. These broadcasts serve as a rich source of socio-cultural data, offering insights into gender dynamics, topic distributions, and potentially emotional patterns in speech. 
To systematically investigate these potential differences, we focused on the following parameters:

\textit{Question 1:} Speaker Total Air Time Duration: Is there a clear difference between male and female speakers in total air time duration on news and talk radio broadcasts?

\textit{Question 2:} Air-Time Allocation at Different Day Times: Does the distribution of air-time for male and female on-air vary during different times of day?

\textit{Question 3:} Topic Representation: Are there pronounced differences between male and female speakers in topic representation?

To answer these questions, we developed a method for automatically collecting and processing radio broadcast data. For data collection, we implemented a multi-channel online radio recorder. The subsequent preprocessing, speaker feature extraction, and model inference stages were carried out using, alongside other models and libraries, our developed framework VANPY, a Python-based multi-model framework for speech analysis and speaker characterization~\cite{koushnir2025vanpyvoiceanalysisframework} (see Figure~\ref{fig:flow}). 
The preprocessing pipeline consists of several steps. First, the raw audio streams are converted to a standardized format 
for consistent processing. Next, the audio undergoes speaker diarization to segment the recordings by individual speakers. For each voice segment, speaker embeddings along with acoustic features are extracted. These embeddings and features are then used at the model inference stage for multiple tasks: audio classification to isolate speech segments from other audio content and speaker characterization for gender classification. The pipeline concludes with speech-to-text transcription, which enables topic analysis through NLP-based zero-shot classification. The list of candidate topics was generated by analyzing speech content using a large language model (LLM). 

To evaluate our method, we conducted an extensive experiment. We collected data from broadcast recordings from 136 online news and talk radio stations across the United States. We selected news/talk radio because it is the most listened-to radio format in the United States~\cite{forbes2024radio}, and collected recordings over a 24-hour period. After filtering out stations with less than 12 hours of talk content, 
our final dataset consists of 74 stations, yielding over 1,400 hours of recorded content for analysis. 
Using our pipeline, we processed this data to generate a large-scale annotated dataset of news and talk radio broadcasts, providing rich metadata about speakers, content, and temporal patterns.

Our analysis reveals consistent patterns of gender disparity in total on-air speaking time, across all investigated stations. Female representation in air-time varies significantly, ranging from 47.5\% in the most balanced station to less than 12\% in the most extreme station, indicating a persistent gender gap even in the most gender balanced cases (see Figure~\ref{fig:gender_extremes}). 
The temporal distribution of gender representation also shows notable patterns throughout the day. Male hosts predominantly lead during prime time slots (morning and evening hours) (see Figure~\ref{fig:time_balanced}). Female presenters tend to have an increased presence during off-peak broadcasting hours, pointing to potential structural bias in time slot allocation.
Moreover, gender-based disparities appear also in topic diversity. Female speakers reached their highest representation in Entertainment News topic segments (35.8\%). Their presence is markedly reduced in all other topics. Particularly striking is the minimal female representation in Politics and Government coverage (15.7\%) as well as talk show segments (10.8\%). Many of these topics are associated with content that may shape public discourse and opinion, particularly news, politics, public safety, health, and community affairs (see Figure~\ref{fig:topic_gender_ordered}).

Our research provides several key contributions to the study of gender representation in broadcast media. First, we introduce a novel computational approach for evaluating gender diversity in voice data at scale, leveraging advanced speech processing techniques and machine learning models. This methodology not only offers a comprehensive analysis of gender representation patterns, but also provides a replicable framework for future studies of broadcast and other types of voice content. Our findings yield quantitative evidence for persistent gender biases in radio broadcasting, with detailed evaluation on how these inequalities are manifested across different temporal and topical dimensions.

The rest of this paper is structured as follows: Section~\ref{sec:rw} reviews related work in three key areas: radio broadcast analysis, gender representation studies, and technical tools for broadcast analysis. Section~\ref{sec:method} describes our methodology and experimental setup, including data collection procedures and the implementation of our analysis framework. Section~\ref{sec:results} presents our findings on gender representation in radio broadcasting, including distributions of speaking time and topic analysis. Section~\ref{sec:dis} discusses the broader implications of our results and the potential applications of our methods. Finally, we conclude with a summary of our contributions and future directions.

\section{Related Work}
\label{sec:rw}
This section presents three main areas of related research. First, we review technological advances in automated analysis of radio broadcasts, where speech recognition and large-scale content processing have enabled the systematic study of talk radio patterns and discourse (see Section~\ref{ssec:radio_broadcas_analysis}). Second, we examine research on gender representation in radio and broader media, showing consistent findings of clear underrepresentation of females  across various platforms and roles (see Section~\ref{ssec:gender_gap_analysis}). Finally, we discuss the evolution of technical tools for broadcast analysis, from individual components like speaker diarization and speech recognition to integrated frameworks that enable comprehensive audio processing and speaker characterization (see Section~\ref{ssec:tools_radio_broadcast}).

\subsection{Automated Analysis of Radio Broadcasts}
\label{ssec:radio_broadcas_analysis}
\subsubsection{Data Collection}
Earlier voice analysis research of radio talk and call-in shows has used mainly manual, qualitative conversation-analytic methods applied to recordings from one or a small number of programmes, and reported a range of distinctive discourse patterns~\cite{hutchby2005media}. Studies exposed features such as standardized opening sequences, power asymmetries in host-caller interactions, and callers' use of ``witnessing" devices to authenticate their contributions \cite{hutchby2005media, hutchby2013confrontation}. While these rule-based analyses provided valuable insights into radio talk dynamics, they were limited to small samples, making automated analysis of large broadcast archives a promising direction for investigating these patterns at scale.

Recent advances in speech analysis technology have enabled large-scale radio content analysis, as demonstrated by the RadioTalk corpus~\cite{Beeferman2019}, which contains 2.8 billion words from 284,000 hours of U.S. talk radio broadcasts. RadioTalk includes speech-to-text (STT) transcripts with timing for each speech fragment, as well as inferred speaker gender, a flag indicating whether an utterance was recorded in a studio or came from a telephone call-in, and program or show identifiers and names collected from scraped station schedules. These annotations were generated using the Kaldi~\cite{povey2011kaldi} and LIUM~\cite{meignier2010lium} toolkits.

\subsubsection{Feature extraction and analysis approaches}
Large-scale radio corpora can include not only transcripts generated by automatic speech recognition (ASR), which alone are accompanied by accuracy limitations (word error rates of 13.1\% for RadioTalk~\cite{Beeferman2019}), but also metadata on speaker characteristics derived directly from the audio signals. 
For example, \citet{FURNER2021115236} introduced a framework for automated analysis of radio broadcast music content that combines machine learning-based audio classification with novel visualization techniques. The system was developed for data collection from online music services to enable automated monitoring of programming (genres, emotions) and station similarity analysis. Similarly, \citet{amrane2022deep} proposed a deep hybrid model for advertisement detection in broadcast TV and radio content, 
combining audio segmentation using silence boundaries, acoustic features extraction, and advertisements detection and classification model inheritance. 

\subsubsection{Applications and examples}
These methodological advances have supported a range of applications in radio research. 
\citet{brannon2020mapping} used a large-scale corpus of transcribed U.S. talk radio broadcasts to show how corporate ownership shape much of the content, while locally produced programming remains relatively limited. 
\citet{FURNER2021115236} evaluated their music-content framework on nine FM radio stations, 
showing improved genre-classification accuracy 
and demonstrated improved accuracy in automated genre classification for radio music content.
\citet{amrane2022deep} reported that their advertisement detection system outperforms available open-source baselines.
More recently,~\citet{mittal2024wavepulse} analyzed the 2024 Presidential Election using WavePulse, a framework for real-time content analytics of radio livestreams. The study monitored 396 news radio stations over three months, processing nearly 500,000 hours of audio streams through speech recognition, speaker diarization, and sentiment analysis. This data was used to track political narratives, analyze content syndication patterns and measure candidate sentiment trends at national and state levels.

\subsection{Gender Representation in Radio and Other Media Domains}
\label{ssec:gender_gap_analysis}
Research on gender representation differences in radio broadcasts and other media channels has been a rising interest over the past decade, revealing both on-air disparities and underlying structural barriers.

For example, in 2017 ~\citet{OBrien18082019} investigated women's participation in community radio in Ireland, revealing significant female underrepresentation. 
In light of organizational and social barriers that hindered female's equal engagement in the sector, the study also raised the need for collective initiatives and structural changes to promote gender equality within community radio organizations.
In 2020,~\citet{crider2020tomatoes} demonstrated a gender gap in on-air time for music artists, finding that of 191 radio stations, only one music format (Top-40) featured a majority of female singers. In the same year,~\citet{10_1093_cdj_bsz030} explored the role of community radio in promoting gendered participation among women in Northern England. 
Furthermore,~\citet{Williamson03072021} provided clear evidence of female underrepresentation on commercial FM radio stations in the top 20 U.S. markets, showing that women make up only about one-third of on-air talent and are especially rare in key roles like lead morning hosts.
More recently, \citet{pelloin2024automatic} investigated gender biases in news subjects across French TV and radio broadcasts using computational methods for transcription and classification, analyzing over 11,700 hours of broadcasts from 21 French channels. Their results demonstrated significant underrepresentation of female speakers in topics such as sports, politics, and conflicts, while showing higher female representation in health, weather, and commercial segments. The study highlights the potential of NLP models to monitor gender biases in media content and suggests that such computational approaches can reveal structural imbalances in representation. 
Another recent study compared representation gaps across TV and radio broadcasts, finding that women's presence in French radio was 41-44\% according to manual reports, while speaking time was lower: 37.9\% on public stations and 28.3\% on private stations \citep{doukhan2024gender}.

Female underrepresentation has also been documented in other media types. For example,~\citet{Kagan2020} demonstrated an existing gender bias in the film industry, despite a trend of improvement in multiple aspects of female roles in movies, including the centrality of characters. Their study introduced a new Gender Ratio test requiring at least 50\% female speaking character, which showed that only 12\% of movies meet this stricter standard for female representation (compared to Bechdel test). Similarly, a year later,~\citet{asr2021gender} showed that men are quoted approximately three times as often as women in online news articles from major Canadian outlets. 
At the global level, the 2025 Global Media Monitoring Project (GMMP), conducted on 6 May 2025, found that women constituted only 26\% of people seen, heard, or spoken about in print and broadcast news overall, and only 22\% in radio news specifically~\cite{GMMP2025Highlights}. These indicators have remained largely stagnant since around 2010.

Taken together, these studies underscore the widespread nature of gender disparities across various media platforms, highlighting the need for continued efforts to measure and address gender inequality in media representation.

\subsection{Tools for Radio Broadcast Analysis}
\label{ssec:tools_radio_broadcast}

Automated analysis of news and talk radio has evolved from ``single-component'' signal-processing pipelines (e.g., standalone diarization or ASR) into integrated systems that can (i) segment continuous streams into speaker-homogeneous speech regions, (ii) represent each region using robust acoustic features and speaker embeddings, (iii) infer speaker attributes (e.g., gender, emotion), and (iv) convert speech into text for content indexing and topic-level analysis. This evolution was enabled by advances in deep learning for audio and speech processing \cite{purwins2019deep} and by the maturation of open toolkits and pretrained models that make large-scale processing feasible.

\paragraph{From early broadcast toolkits to modern end-to-end pipelines.}
A useful historical reference point is the creation of large-scale talk-radio corpora such as RadioTalk, whose transcripts and metadata were generated using classical ASR/diarization toolkits including Kaldi and LIUM \cite{Beeferman2019,povey2011kaldi,meignier2010lium}. More recent broadcast analytics systems move toward near-real-time, multi-stage pipelines that combine diarization, ASR, and downstream NLP tasks; for example, WavePulse monitors hundreds of stations over long periods and applies ASR, diarization, and sentiment/content analytics at scale \cite{mittal2024wavepulse}. In parallel, computational media studies on broadcast TV/radio increasingly rely on these pipelines to quantify representation patterns and topic biases using transcription plus automatic classification \cite{pelloin2024automatic,doukhan2024gender}.

\paragraph{Segmentation and diarization (``who speaks when'').}
At the audio level, the first critical step is separating speech from non-speech and partitioning speech into speaker-homogeneous segments. Modern diarization approaches are typically deep-learning based and can be used as modular building blocks in larger pipelines. In our workflow we rely on \texttt{pyannote.audio} \cite{bredin2020pyannote}, which provides neural components and optimized pipelines for speaker segmentation, clustering, and speaker-turn assignment. This step is particularly important in talk radio, where studio speech, phone call-ins, and background audio can vary widely in quality and channel characteristics.

\paragraph{Acoustic features and speaker representations.} 
After segmentation, radio analysis pipelines commonly extract both (a) handcrafted acoustic descriptors and (b) learned speaker embeddings. Handcrafted features such as MFCCs, energy- and spectrum-based measures, and temporal statistics remain widely used due to interpretability and efficiency, and are often computed with established digital signal processing libraries (e.g., \texttt{librosa}) \cite{brian_mcfee_2022_6097378}. However, for robust speaker characterization under channel and noise variability, modern systems frequently rely on embedding representations (e.g., x-vectors / related deep speaker representations) produced by pretrained neural models and toolkits such as \texttt{SpeechBrain} \cite{speechbrain}. In our pipeline, embeddings are generated with an ECAPA-TDNN-based extractor (as implemented in the used toolkit) \cite{sheikh2022introducing}, and combined with conventional acoustic features for downstream inference.

\paragraph{Audio event filtering and speech-to-text (``what is being said'').}
Broadcast recordings contain non-speech events (music beds, jingles, advertisements, noise) that can harm both diarization and ASR. Therefore, many pipelines include an audio event classifier to filter or annotate non-speech regions. In our system we use YAMNet \cite{Plakal_Ellis} to support this filtering step. For transcription, current broadcast-oriented studies increasingly adopt large pretrained ASR models that are robust to real-world acoustic conditions. In particular, Whisper \cite{radford2022whisper} is widely used as a strong baseline for transcription of broadcast speech and is an enabling component for downstream content indexing and semantic analysis; recent broadcast-media studies similarly rely on transcription followed by topic-oriented classification to quantify representation and content patterns \cite{pelloin2024automatic,doukhan2024gender,mittal2024wavepulse}.

\paragraph{From transcripts to topic models (NLP layer).}
Once speech is transcribed, the analysis typically moves to an NLP layer where utterances (or aggregated segments) are assigned topics, frames, or discourse categories. A practical family of tools here is ``zero-shot'' text classification using pretrained NLI models (e.g., BART-based models), which can be applied when labeled training data is limited and topic taxonomies evolve over time \cite{lewis2019bart}. In addition, recent pipelines often use large language models to assist with taxonomy construction or topic-list generation, which can then be operationalized via a deterministic classifier in the processing stage \cite{anthropic2023claude}.

\paragraph{Integrated frameworks and reproducibility.}
Together, these developments show a shift from isolated speech-processing components toward integrated, reproducible pipelines for large-scale broadcast analysis. In this study, we build on this direction by applying VANPY, a Python-based voice analysis framework, as the execution backbone for preprocessing, feature extraction, speaker characterization, transcription, and downstream topic analysis~\cite{koushnir2025vanpyvoiceanalysisframework}. The specific implementation of this pipeline is described in Section~\ref{sec:method}.

\section{Methods and Experiments}
\label{sec:method}
The primary objective of this study is to investigate gender related speaker and content trends in news and talk radio broadcasts. Specifically, the research focuses on speech duration, time of day representation, and topic distributions. 
The processing workflow was implemented using VANPY, a Python-based voice analysis framework designed for automated preprocessing, feature extraction, and speaker characterization~\cite{koushnir2025vanpyvoiceanalysisframework}. VANPY organizes processing into a sequence of stages, including audio preprocessing, feature extraction, and model inference, connected through an inter-component ``payload.'' 
In this study, VANPY was used as the execution backbone for integrating diarization, audio classification, speaker embedding extraction, acoustic feature extraction, gender classification, and speech-to-text transcription.
To achieve these objectives, data was collected using a custom-developed online-radio stream recorder capable of simultaneously recording multiple stations. The collected recordings were then processed using the VANPY-based workflow described below.

\subsection{Data Collection}
The evaluation dataset was created by recording online radio broadcasts, focusing exclusively on news and talk stations to emphasize natural speech and minimize music content. The list of stations was derived from the Radio Station USA Talk Radio directory,\footnote{\url{https://radiostationusa.fm/category/talk}} which categorizes radio stations by genre. For each station, an attempt was made to locate it in the RadioGarden API~\cite{radio_garden}, and the content was manually verified to ensure it was primarily news and talk-oriented and correctly placed in its respective state. This process accounted for cases where stations with the same name had different content or locations across states.

Using this curated list, radio stations applicable to recording via the Radio Garden API were recorded simultaneously over a 24-hour period, from 2024-07-31 23:00:00 UTC to 2024-08-01 23:00:00 UTC. Recordings were segmented into 10-minute chunks to facilitate processing and minimize data loss in case of a network or other types of errors. After the recording, stations with less than 12 hours of speech content were filtered out to ensure sufficient data quality and consistency.

As a result, the dataset includes 74 stations, each contributing between 12 and 19.6 hours of speech content, amounting to more than 1,400 hours of speech in total (175 GB). This dataset serves as the foundation for analyzing speaker characteristics and content trends across U.S. news and talk radio broadcasts.

\subsection{Framework and Third-Party Model Execution}
\begin{figure}[!htb]
 \centering
 \captionsetup{justification=centering}
 \makebox[\textwidth]{\includegraphics[width=1\textwidth]{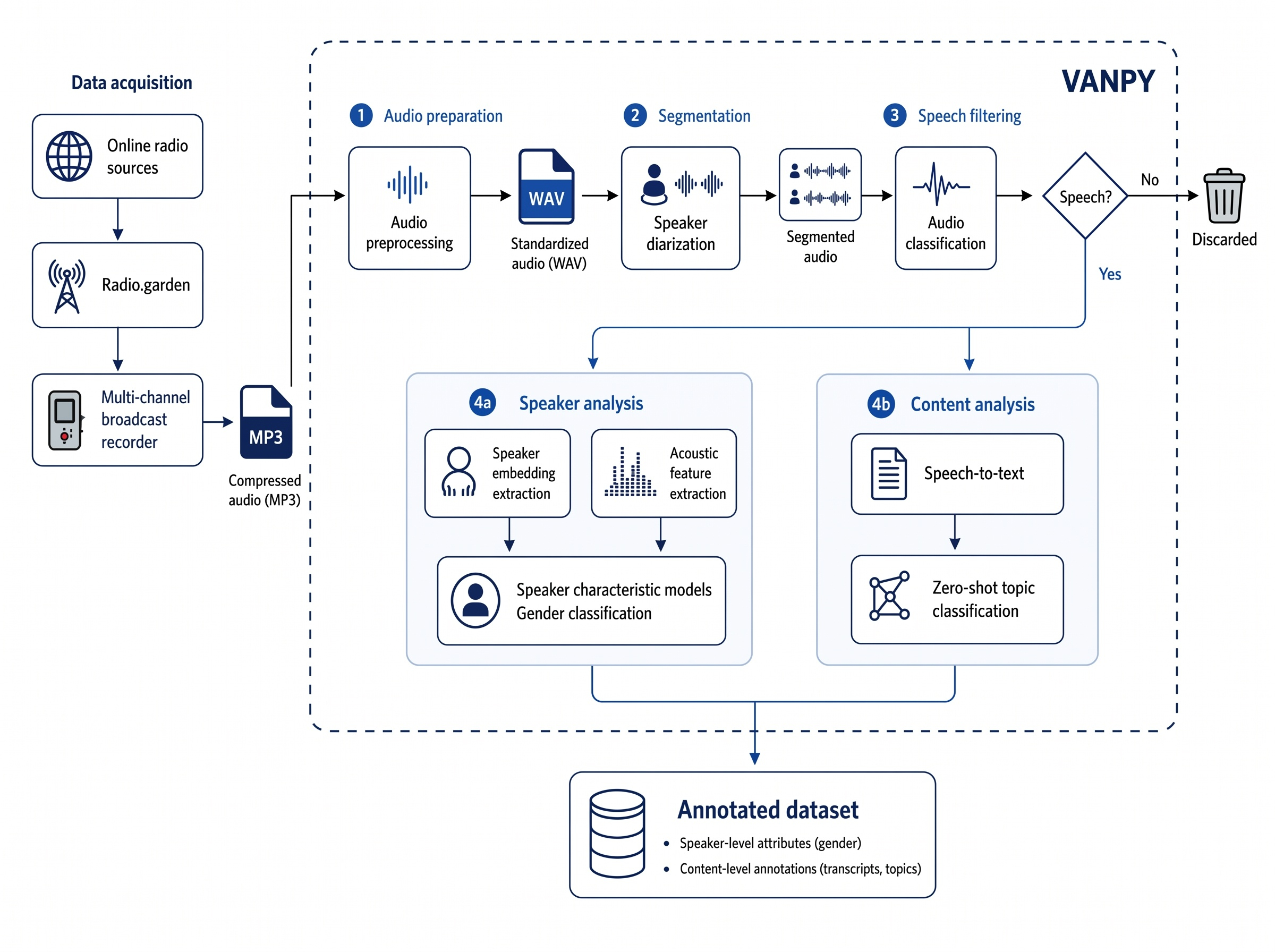}}
  \caption{Processing pipeline for radio broadcast analysis: from broadcast streams to annotated dataset generation, including speaker diarization, embedding extraction, audio classification, speaker characterization, and content analysis.}
  \label{fig:flow}
\end{figure}

The processing workflow applied to the collected recordings is illustrated in Figure~\ref{fig:flow}. First, each recording was standardized into a common audio format. Speaker diarization was then used to divide the recordings into speaker-homogeneous segments. For each segment, we extracted speaker embeddings and acoustic features, applied speaker gender classification, and generated speech-to-text transcriptions. Finally, the transcribed text was used for topic classification.

\begin{enumerate}
    \item \textbf{Audio Preprocessing}: Each audio file was converted into an uncompressed WAV format with standardized parameters: single-channel, 256 kbps bitrate, and 16 kHz sampling rate.
    \textit{Input}: Compressed MP3 files
    \textit{Output}: Resampled WAV files compatible with downstream models
    
    \item \textbf{Speaker Diarization}: Pyannote speaker diarization~\cite{bredin2020pyannote} was used to segment and cluster audio data by speaker identity, incorporating a Voice Activity Detection (VAD) unit for precise segmentation.
    \textit{Input}: Full-length WAV files
    \textit{Output}: Speaker-segmented WAV files with unique speaker ID labels
    
    \item \textbf{Audio classification}: The YAMNet classifier~\cite{Plakal_Ellis} categorized audio events across 521 classes to identify non-speech audio cues.
    \textit{Input}: Speaker-labeled audio segments
    \textit{Output}: Content classification labels (primarily filtering for "speech" segments)
    
    \item \textbf{Speaker Embedding}: The SpeechBrain x-vector speaker embedding extractor~\cite{speechbrain} was applied to generate robust speaker embeddings using ECAPA Time-Delay Neural Network (ECAPA-TDNN).
    \textit{Input}: Speech WAV segments
    \textit{Output}: 192-dimensional speaker latent representations
    
    \item \textbf{Acoustic Feature Extraction}: The Librosa library~\cite{brian_mcfee_2022_6097378} was utilized to extract features relevant for specific tasks, such as Mel-Frequency Cepstral Coefficients (MFCC), delta-MFCC, Zero-Crossing-Rate, Spectral Centroid, Spectral Bandwidth, Spectral Contrast, and Spectral Flatness.
    \textit{Input}: Speech WAV segments
    \textit{Output}: Multi-dimensional acoustic feature vectors
    
    \item \textbf{Speaker Characterization}: We employed the VANPY Gender Classification model from the VANPY framework~\cite{koushnir2025vanpyvoiceanalysisframework}. This model is a robust gender classifier that achieves high accuracy across multiple datasets, allowing us to reliably identify speaker gender throughout the radio broadcast speech segments.
    \textit{Input}: Speaker embeddings from previous steps
    \textit{Output}: Operational binary gender classification (male/female) based on vocal characteristics

    \item \textbf{Speech-to-Text (STT) Transcription}: The OpenAI Whisper large model~\cite{radford2022whisper} was employed for high-quality transcription of spoken content.
    \textit{Input}: Speech WAV segments
    \textit{Output}: Transcribed text content
\end{enumerate}

The accuracy of the gender classification was manually tested on 370 randomly selected segments, stratified by \textit{channel\_id} with 5 samples per channel. The observed accuracy was 97.8\%, with only 8 samples showing classification issues. Among these problematic cases: two samples had speakers whose gender listeners couldn't reliably determine (one due to storage issues, the other was ambiguous despite adequate duration); three samples contained mixed speech from both men and women (indicating segmentation problems); one sample showed a female misclassification due to degraded call-in audio quality; and the remaining samples were a male misclassification affected by significant background music.

After obtaining the STT transcription of the voice content, we applied Natural Language Processing (NLP) models to extract information about the topics discussed in each segment. We only used segments that contained at least 10 words in transcription. For topic classification, we used the \textit{facebook/bart-large-mnli}\footnote{\url{https://huggingface.co/facebook/bart-large-mnli}} zero-shot text classification model. This model requires a predefined list of candidate labels during inference. To generate this list, we utilized Claude 3.5 Sonnet\footnote{by Anthropic, https://claude.ai/, claude-3-5-sonnet-20241022} by providing it with the transcription of a 24-hour recording from one station. Claude 3.5 Sonnet generated 20 inclusive and general topics (see Appendix~\ref{topic_list}).

The accuracy of the topic classification was manually tested on 222 randomly selected segments, stratified on \textit{channel\_id}, 3 samples per channel. Compared with manual annotations, the classifier achieved an overall accuracy of 46\%, and a weighted F1 score of 0.49 across 20 candidate classes. The main sources of classification error are discussed in Section~\ref{sec:dis}.

This workflow allowed for a detailed analysis of speaker gender representation and topic distributions.
The general method flow of the VANPY components is shown in the dashed rectangle in Figure~\ref{fig:flow}.

\section{Results}
\label{sec:results}

\subsection{Male vs Female Air Time Duration}
\begin{figure}[!htb]
 \centering
 \captionsetup{justification=centering}
 \makebox[\textwidth]{\includegraphics[width=\textwidth]{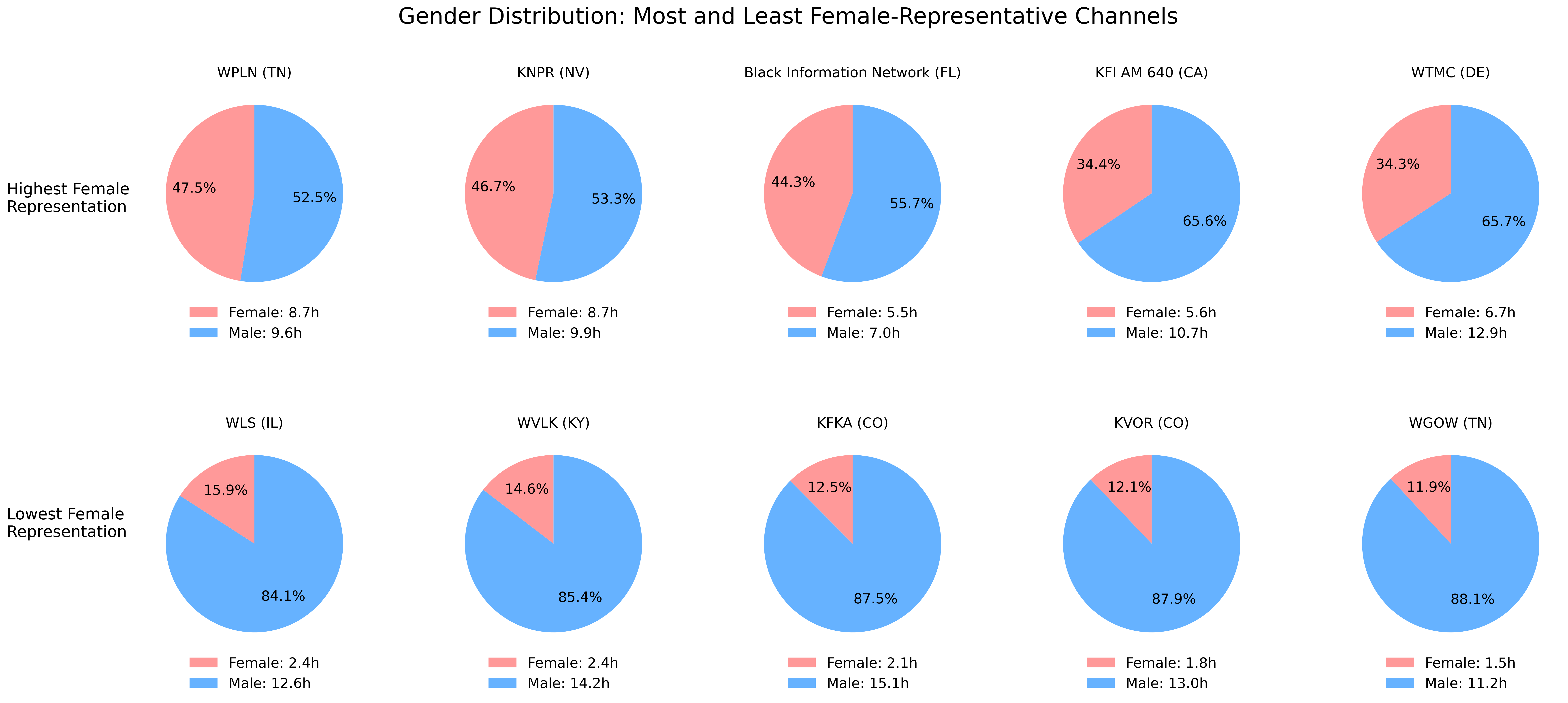}}
  \caption{Gender representation extremes in radio broadcasting (out of 74 stations): Top row shows five stations with highest female representation, the bottom row shows five stations with lowest female representation}
  \label{fig:gender_extremes}
\end{figure} 

Analysis of speaking time distribution across 74 radio stations revealed substantial gender disparity in news and talk radio programming. The average female speaking time across all stations was 23\% (SD = 6.8\%), with male speakers dominating at 77\% of the total airtime. 

Figure~\ref{fig:gender_extremes} illustrates the five stations with the highest and lowest female representation through pie charts. The results show that speech time for these stations ranges from 12.5 to 19.6 hours. The highest female representation was observed at the WPLN (TN) station with 47.5\%, followed by KNPR (NV) at 46.7\%, and the Black Information Network (FL) at 44.3\%. In contrast, several stations demonstrated notably low female representation rates, with KFKA (CO) at 12.5\%, KVOR (CO) at 12.1\%, and WGOW (TN) at 11.9\%. 
Across the full set of 74 stations, only eight stations (10.8\%) had female speaking-time representation above 30\%, while twenty-eight stations (37.8\%) had female representation below 20\%. Public or non-commercial stations such as WPLN (TN), KNPR (NV), and WCSP (DC), as well as the Black Information Network (FL), tended to show relatively higher female speaking-time representation compared with the rest of the dataset. Excluding these relatively more balanced stations, the remaining stations had a mean female speaking-time representation of 22.6\%.

\subsection{Daily Speaking Time Distribution by Gender}
\begin{figure}[H]
 \centering
 \captionsetup{justification=centering}
 \makebox[\textwidth]{\includegraphics[width=\textwidth]{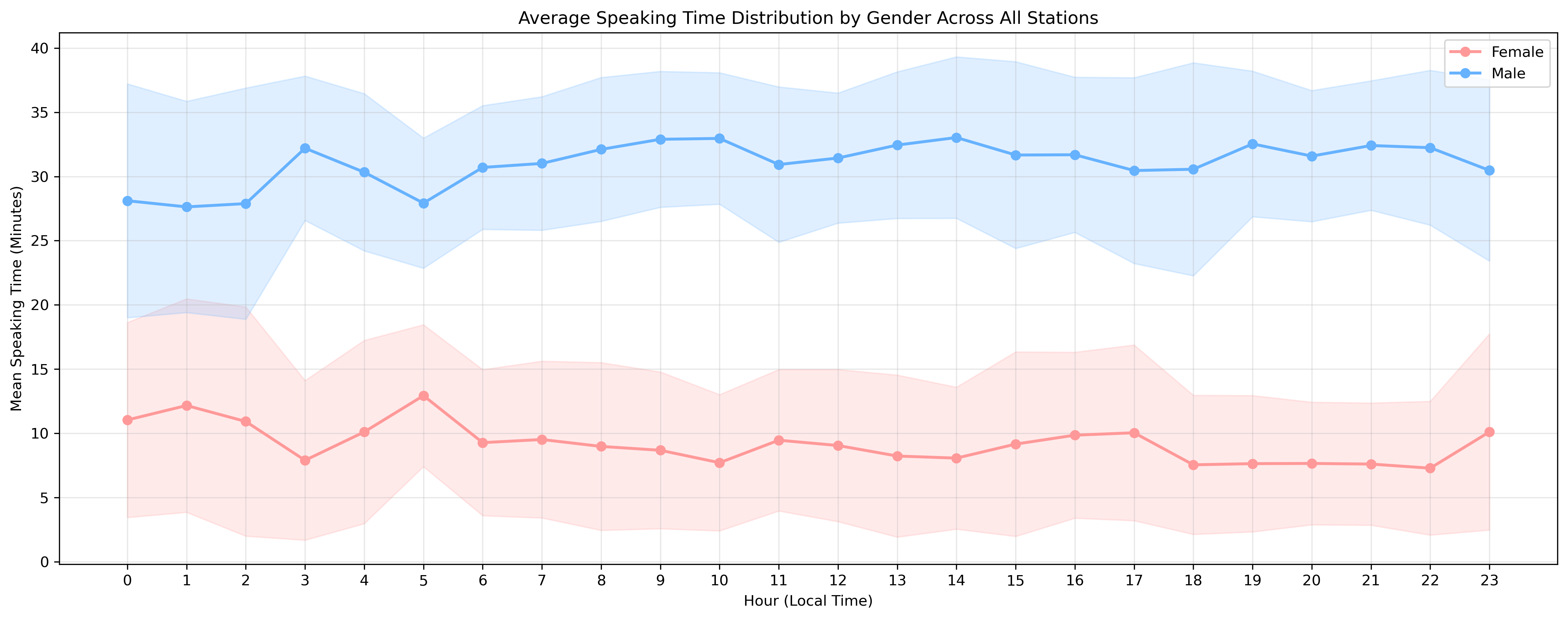}}
  \caption{Mean and standard deviation of speech time over the day, per gender}
  \label{fig:gender_means_hr}
\end{figure}

\begin{figure}[H]
 \centering
 \captionsetup{justification=centering}
 \makebox[\textwidth]{\includegraphics[width=\textwidth]{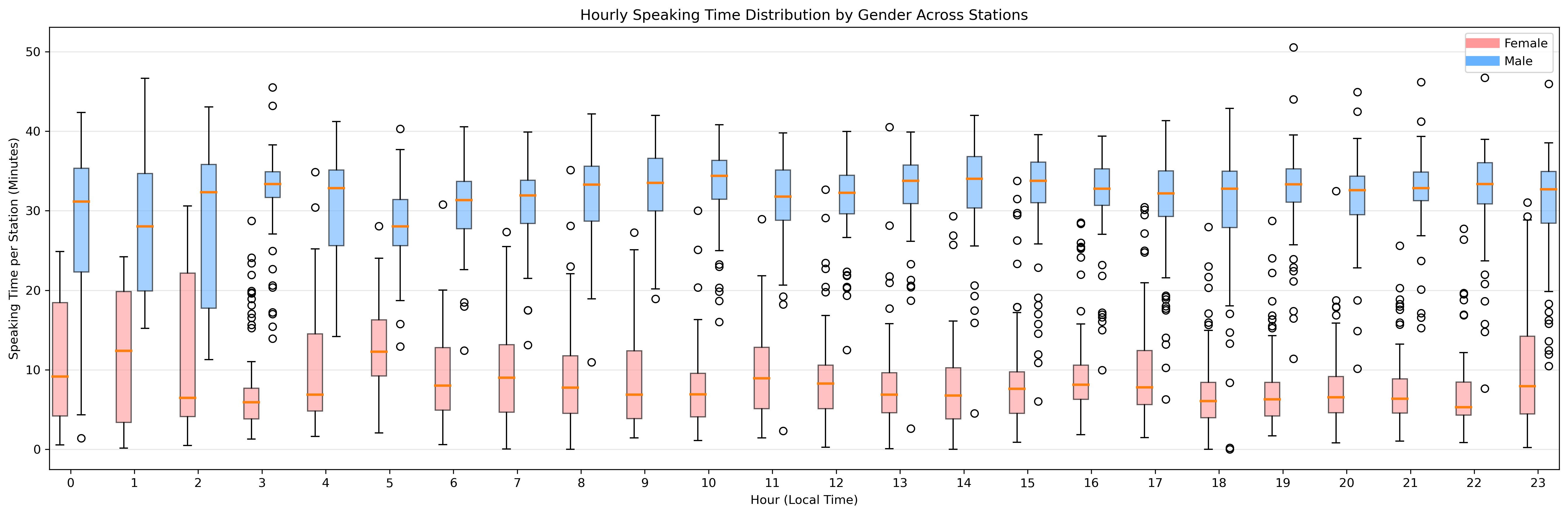}}
  \caption{Hourly distribution of speaking time across stations by gender. Each box represents the station-level distribution for a given hour, with the median, interquartile range, whiskers, and outliers shown separately for female and male speakers}
  \label{fig:gender_bp_hr}
\end{figure}

\begin{figure}[H]
 \centering
 \captionsetup{justification=centering}
 \makebox[\textwidth]{\includegraphics[width=\textwidth]{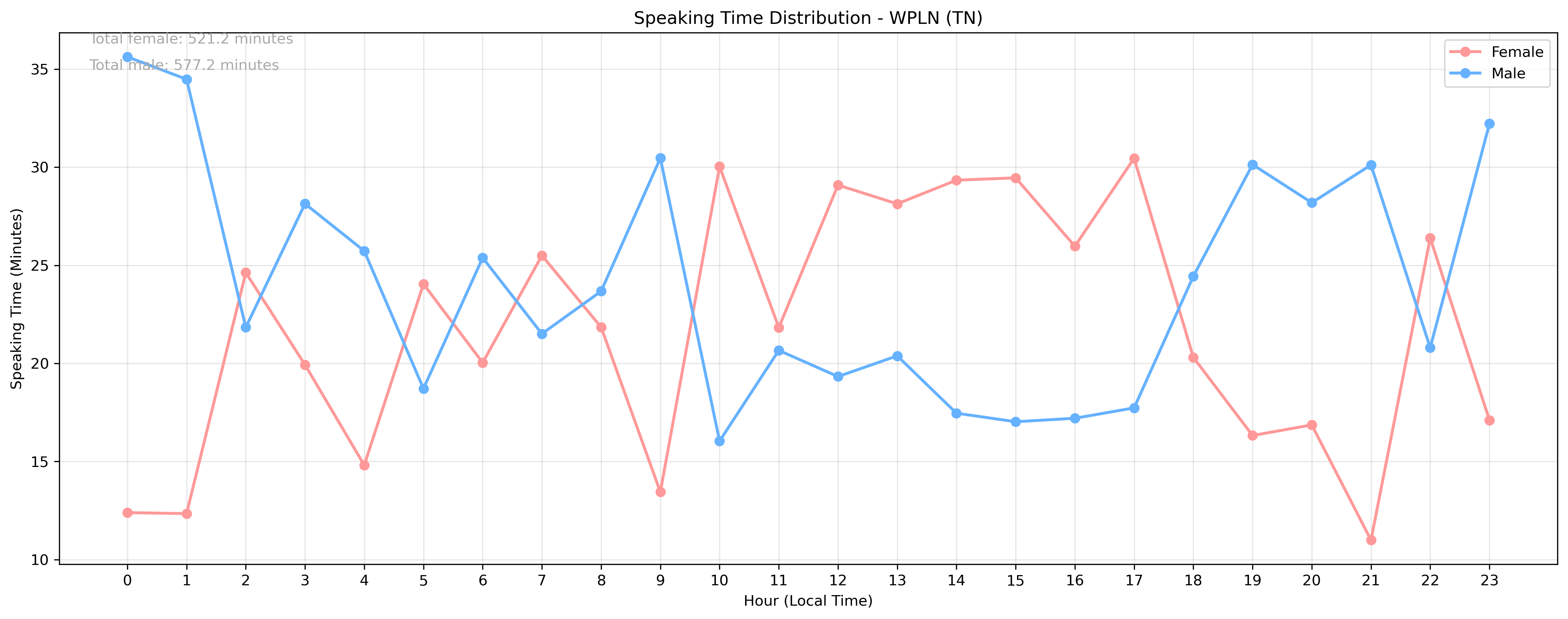}}
  \caption{24-hour speaking time distribution in a gender-balanced station, showing near-equal representation with 47.5\% female speaking time}
  \label{fig:time_balanced}
\end{figure}

Figure~\ref{fig:gender_means_hr} shows the mean and standard deviation of male/female speaking times over all stations by hour. On average, male speakers accounted for substantially more airtime than female speakers throughout the day, with male speaking time generally around 30-33 minutes per hour compared with approximately 8-12 minutes per hour for female speakers. The gap remained visible across nearly all hours, indicating that the underrepresentation of female speakers was not limited to a specific time of day. Female speaking time ranged from 7.3 minutes at 10:00 PM to 12.9 minutes at 5:00 AM, while male speaking time remained consistently higher across the 24-hour period.

Figure~\ref{fig:gender_bp_hr} further examines the station-level distribution behind these hourly averages. For every hour, the median male speaking time across stations was higher than the median female speaking time, and the male interquartile ranges were generally positioned above the corresponding female interquartile ranges. This indicates that the gender gap was not driven only by a small number of highly male-dominated stations, but was reflected in the typical hourly distribution across stations. At the same time, the presence of female outliers and occasional higher female values shows that some stations or specific time slots did reach higher female representation. However, these cases were not representative of the overall distribution.

Figure~\ref{fig:time_balanced} depicts the total airtime duration on WPLN (TN), a Nashville public radio/NPR news and talk station and the most gender balanced station found on this study, showing near-equal gender representation. On this station, in fact, during midday hours (10:00 AM-5:00 PM), women spoke more than men, reaching up to 31 minutes per hour. However, this pattern reversed during the evening hours, including the 6:00-8:00 PM period, when male speaking time again exceeded female speaking time. This suggests that even in the most balanced station, gender representation may vary substantially by program schedule and daypart.

\subsection{Topic Distribution by Gender}
\begin{figure}[H]
 \centering
 \captionsetup{justification=centering}
 \makebox[\textwidth]{\includegraphics[width=\textwidth]{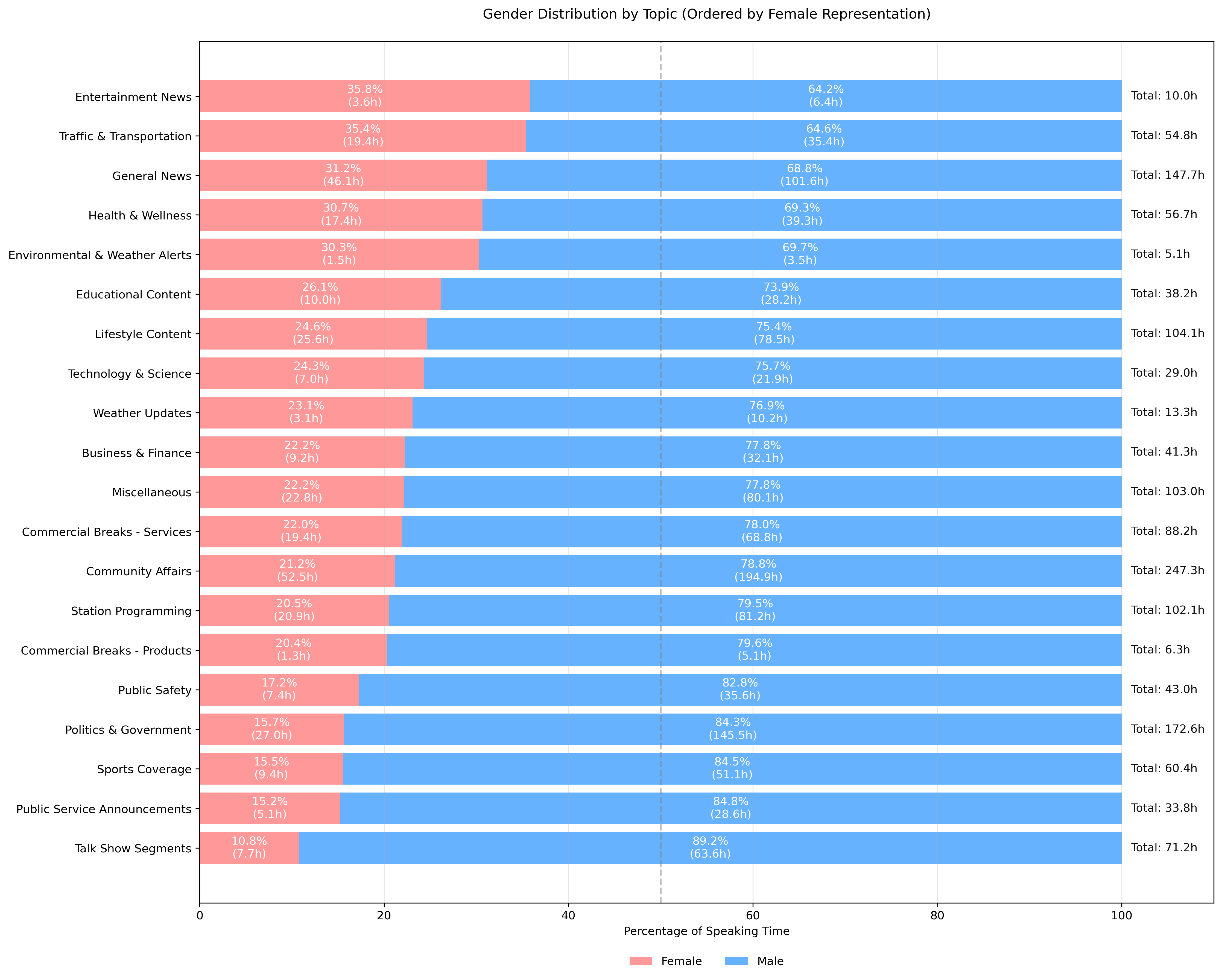}}
  \caption{Gender Distribution by Topic (Ordered by Female Representation)}
  \label{fig:topic_gender_ordered}
\end{figure}

Figure~\ref{fig:topic_gender_ordered} depicts the distribution of speaking time across topics by gender, across all 74 analyzed stations. A clear persistent male dominance can be seen across all content categories.
Community Affairs was the most talked-about category overall (about 247 total hours), with male speakers taking around four times as much talk time as female speakers. Politics \& Government came in second (172 total hours), showing an even stronger male focus (about 146 hours male vs. 27 hours female). General News placed third (148 total hours), where females spoke about 46 hours compared to 102 for males.
Entertainment News showed the highest female representation (35.8\%), followed by Traffic \& Transportation (35.4\%) and General News (31.2\%). The most underrepresented categories, such as Public Service Announcements and talk show segments had lower female talk duration percentages, of 15.2\% and 10.8\%, respectively.

\section{Discussion}
\label{sec:dis}
In this study we analyzed gender differences of speaking representation in radio broadcasting across 74 U.S. news and talk radio stations, using filtered recordings collected over a 24-hour period and totaling more than 1,400 hours of content.
Our analysis revealed consistent patterns of gender disparity in total on-air speaking time, daily time slots, and content topics. These findings extend previous studies showing female underrepresentation in community radio participation \citep{OBrien18082019,10_1093_cdj_bsz030}, music radio airtime \citep{crider2020tomatoes}, and commercial FM on-air talent and hosting roles \citep{Williamson03072021}. The main contribution of our study is not merely to show that a gender gap exists, but to characterize how this gap appears in U.S. news and talk radio when measured directly from broadcast audio. Specifically, our large-scale audio-based analysis quantifies male and female speaking time at the station level and shows how the imbalance varies across the broadcast day and across content topics. This provides evidence that underrepresentation is reflected not only in participation or staffing patterns, but also in listeners’ actual exposure to male and female voices, including during peak listening hours and in public-discourse topics such as politics, community affairs, and talk show segments.
Overall, male speakers accounted for 77\% of the total airtime. 

The station-level distribution further indicates that the observed gender gap was not driven by a small number of highly male-overrepresentative stations, but reflected a broader pattern across the dataset. Although several public, non-commercial, or diversity-oriented stations showed relatively more balanced representation, they remained exceptions rather than the norm. This distinction suggests that institutional context and programming orientation may be associated with differences in female speaking-time representation, but even the most balanced stations did not fully eliminate the gap.
Importantly, prior research identified underlying barriers to women's equal participation in radio, including gendered violence, caregiving responsibilities, and technical fears that limit involvement in community radio~\cite{10_1093_cdj_bsz030}. 

Temporal analysis of speaking time, averaged over all stations, demonstrated clear differences between male and female representation throughout the day. Standard-deviation bands overlap only between 1:00-2:00 AM while showing consistent differences throughout all other hours, with male representation dominating at all times of day. Women had relatively greater speaking time early morning around 5:00 AM and at 1:00 AM.
Importantly, female representation remained consistently low throughout the day, including during morning and evening commute hours (6:00-9:00 AM and 6:00-8:00 PM), when radio listening is typically expected to be high.
Male speakers predominantly lead during prime time slots (morning and evening hours). Female speaking time was somewhat higher during selected off-peak hours, pointing to potential structural bias in time slot allocation.
Although the most female-representative station, WPLN (TN), demonstrated a relatively balanced gender ratio, we still observed a drop in the female speaking ratio during evening commute hours (6:00-8:00 PM). Interestingly, at WPLN (TN), men took most of the night shift, contrary to the overall statistics across stations. 

Moreover, gender-based disparities were also evident in topic-level speaking time. Across the 20 content categories, male speakers dominated every topic, although the magnitude of the gap varied substantially. Female representation reached its highest level in Entertainment News (35.8\%), but remained below parity even in this most balanced category. The largest disparities appeared in Talk Show Segments, where women accounted for only 10.8\% of nearly 71 hours of topic-labeled speech, and in Sports Coverage, where female speaking time was approximately 15.5\%. Politics \& Government also showed strong male dominance, with women accounting for only 15.7\% of speaking time.

These differences are particularly important because several of the most discussed categories, including Community Affairs, Politics \& Government, and General News, correspond to public-facing informational content that may contribute to shaping public discourse and opinion. Community Affairs was the most discussed category overall, yet male speakers received approximately four times more airtime than female speakers. Politics \& Government ranked second in total speaking time and showed an even larger gap, with male airtime approximately six times higher than female airtime. General News, the third-largest category by total airtime, was also strongly imbalanced, with 46 hours of female speech compared with 101 hours of male speech. Overall, male-to-female speaking-time ratios ranged from roughly 1.8:1 to 8.3:1 across topics, indicating that gender imbalance was present throughout the topic taxonomy rather than being confined to a small subset of content categories.

We offer the in-house developed analysis VANPY framework as a publicly available package with advanced AI processing for speaker diarization, gender classification, and topic analysis.that could be used for voice analysis in multiple settings and research directions.
This  comprehensive analysis approach for evaluating gender diversity in voice data at scale, leverages advanced speech processing techniques and machine learning models and provides a replicable framework for future studies of broadcast and other types of voice content. 
In this respect, it should be noted that the performance of VANPY's gender classification model was robust, achieving good results despite voice quality issues in call-ins and interference from background music in advertisements. However, despite the effective gender classifier, we found that speaker segmentation presented challenges, especially when multiple voices overlapped in a single segment.

In addition, several methodological choices could introduce biases, such as the minimum word count threshold in transcriptions. We observed that topics were sometimes incorrectly predicted when the complete context wasn't captured within a single segment, or when certain marker words were present. Most classification errors occurred when short segments could reasonably fit under several broad labels, especially with topics like "Community Affairs," "Politics \& Government," "Miscellaneous," and "General News," which were often confused with each other. Advertisements were another source of classification problems. For example, when a drug advertisement included phrases like "for orders call [telephone number]," it was labeled as a "Commercial break," but when discussing the drug's effects, the system classified it as "Health \& Wellness" content. Similarly,  some segments containing weather-related terms were classified as "Weather Updates" even when weather was not the main topic. Future work should explore methods for combining multiple speaking turns into coherent content windows, even when interrupted by pauses or other speakers.

\section{Conclusions}
\label{sec:con}
We performed an automated speaker analysis on online radio broadcasts, revealing substantial gender disparities in multiple aspects of speaking time across news and talk radio. Our analysis of 74 US radio stations demonstrated that male speakers dominated 77\% of total airtime, with additional differences exhibited in time slot allocations throughout the day and gender differences in participation at different topics, with female being significantly underrepresented in prime time and in topics that are most talked about and that have an important role in shaping public opinion.

The methodology developed in this study is publicly shared and can be applied to other audio data sources (e.g., podcasts, recorded conversations, or video soundtracks) to segment content and extract both speaker identity and characteristics (such as gender, age, accent, and emotion). Future work will explore additional speaker characteristics from radio broadcasts and other voice data sources, including emotion, ethnicity, and accent classification to provide a more comprehensive understanding of representation in broadcast media.

\begin{appendices}
\section{A list of examined radio stations}\label{station_list}
\begin{table}[H]
\caption{Recorded Stations by States}
\label{table_recorded_stations}
\centering
\begin{tabular}{cccccccc}
\hline
Station & State & Station & State & Station & State & Station & State \\ \hline
WLOB & ME & WIOD & FL & WAAX & AL & WLVL & NY \\ \hline
WGMD & DE & WBAP & TX & KMJ & CA & WZFG & ND \\ \hline
WRMN & IL & WTAM & OH & WSYR & NY & WFLA & FL \\ \hline
KNEB & NE & WHKY & NC & WAIM & SC & WDWS & IL \\ \hline
KFAB & NE & WMAN & OH & KLIK & MO & WFNC & NC \\ \hline
WHO 1040 & IA & WVLK & KY & WXLM & CT & CKLW & ON \\ \hline
KTRF & MN & WWTC & MN & KFAY & AR & KCOL & CO \\ \hline
WIBC & IN & WMAC & GA & WDAY & ND & WCSP & DC \\ \hline
KFYR & ND & WMAL & DC & KHVH & HI & KQNT & WA \\ \hline
KTBB & TX & KFI AM 640 & CA & KNPR & NV & KPRL & CA \\ \hline
WVHU & WV & KVOR & CO & KDAL & MN & WTMC & DE \\ \hline
WJPF & IL & WCBM & MD & WRAK & PA & KIRO & WA \\ \hline
Black Information Network & FL & WGOW & TN & KTAR & AZ & WERC & AL \\ \hline
WGAW & MA & WMOH & OH & WFTL & FL & WBAL & MD \\ \hline
WPTI & NC & WAAM & MI & KBOI & ID & 710 WOR & NY \\ \hline
WPLN & TN & WGY & NY & WLS & IL & KFKA & CO \\ \hline
KCSJ & CO & WJR & MI & WNIS & VA & KWNO & MN \\ \hline
KAOI & HI & KRDO & CO & WENG & FL & WWNC & NC \\ \hline
KNSI & MN & KKOH & NV &  &  &  &  \\ \hline

\end{tabular}
\end{table}

\begin{figure}[H]
 \centering
 \captionsetup{justification=centering}
 \makebox[\textwidth]{\includegraphics[width=\textwidth]{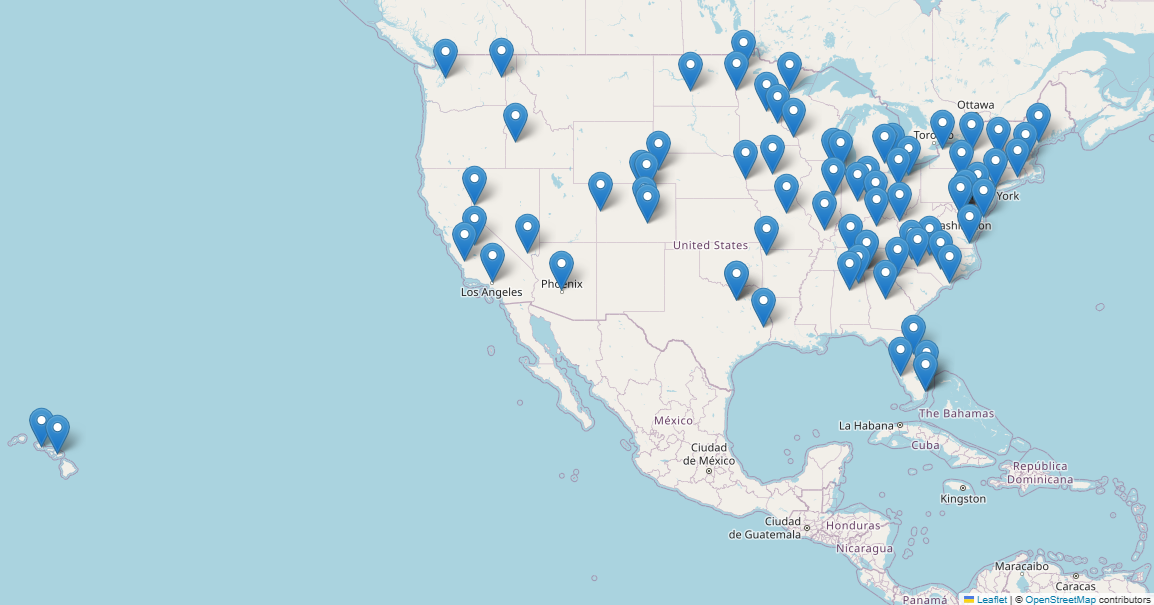}}
  \caption{Geo-locations of the recorded stations}
  \label{fig:geo-stations}
\end{figure}

\section{A list of candidate topics}\label{topic_list}

The topics list was generated using the following prompts:
\begin{enumerate}
    \item I have a PD series of radio recording transcriptions. I want you to extract 49 topics (+ one misc) out of the transcription. Give me the topics based on the attached document
    \item Shrink by generalization the list of topics to at most 20 topics
\end{enumerate}
\begin{table}[H]
\centering
\caption{Candidate Topics}
\label{tab:candidate_topics}
\small 
\begin{tabularx}{\textwidth}{@{}>{\raggedright\arraybackslash}X >{\raggedright\arraybackslash}X@{}}
\toprule
\textbf{Topic} & \textbf{Category} \\ 
\midrule
Radio - General News & Breaking, Local, National, International news \\ 
Radio - Politics \& Government & Political news, commentary, government affairs \\ 
Radio - Business \& Finance & Business news, financial advice, market updates \\ 
Radio - Weather Updates & All weather-related content \\ 
Radio - Traffic \& Transportation & Traffic updates, road conditions, travel info \\ 
Radio - Sports Coverage & All sports-related content, local and national \\ 
Radio - Entertainment News & Music, celebrities, movies, TV \\ 
Radio - Health \& Wellness & Health news, medical advice, wellness tips \\ 
Radio - Technology \& Science & Tech news, scientific developments \\ 
Radio - Community Affairs & Local events, community news, education \\ 
Radio - Public Safety & Emergency alerts, crime reports, public safety info \\ 
Radio - Commercial Breaks - Services & Financial, healthcare, professional services ads \\ 
Radio - Commercial Breaks - Products & Consumer goods, retail, automotive ads \\ 
Radio - Talk Show Segments & Call-ins, interviews, discussions \\ 
Radio - Lifestyle Content & Food, home, garden, personal advice \\ 
Radio - Station Programming & IDs, schedules, promos \\ 
Radio - Public Service Announcements & Community services, public information \\ 
Radio - Educational Content & Learning segments, informative programs \\ 
Radio - Environmental \& Weather Alerts & Severe weather, environmental news \\ 
Radio - Miscellaneous & Other content that doesn't fit above categories \\ 
\bottomrule
\end{tabularx}
\end{table}

\end{appendices}

\bibliographystyle{unsrtnat}
\bibliography{sample-bibliography}

@misc{radio_topic,
	title = {U.S. Radio Industry - Statistics \& Facts},
	howpublished = {\url{https://www.statista.com/topics/1330/radio/#dossierKeyfigures}},    
	author = {{Statista Research Department}},
	note = {[Online; accessed 07-August-2023]},
    year={2009}
}

@article{jauert1997local,
  title={Local radio in Western Europe: Conflicts between the cultures of center and periphery},
  author={Jauert, Per},
  journal={Nordicom Review},
  volume={18},
  number={1},
  pages={93--106},
  year={1997},
  publisher={Nordicom, University of Gothenburg}
}

@article{barker2000political,
  title={Political talk radio and public opinion},
  author={Barker, David and Knight, Kathleen},
  journal={Public opinion quarterly},
  volume={64},
  number={2},
  pages={149--170},
  year={2000},
  publisher={Oxford University Press}
}

@article{doi:10.1016/j.brq.2014.06.001,
author = {Josefa D. Martín-Santana and Clara Muela-Molina and Eva Reinares-Lara and Miriam Rodríguez-Guerra},
title ={Effectiveness of Radio Spokesperson'S Gender, Vocal Pitch and Accent and the Use of Music in Radio Advertising},
journal = {BRQ Business Research Quarterly},
volume = {18},
number = {3},
pages = {143-160},
year = {2015},
doi = {10.1016/j.brq.2014.06.001},

URL = { 
        https://doi.org/10.1016/j.brq.2014.06.001
    
},
eprint = { 
        https://doi.org/10.1016/j.brq.2014.06.001
    
}
}

@article{hurtz2004effects,
  title={The Effects of Gender-Stereotyped Radio Commercials 1},
  author={Hurtz, Wilhelm and Durkin, Kevin},
  journal={Journal of applied social psychology},
  volume={34},
  number={9},
  pages={1974--1992},
  year={2004},
  publisher={Wiley Online Library}
}

@Inbook{Grekow2018,
author="Grekow, Jacek",
title="Emotion Tracking of Radio Station Broadcasts",
bookTitle="From Content-based Music Emotion Recognition to Emotion Maps of Musical Pieces",
year="2018",
publisher="Springer International Publishing",
address="Cham",
pages="85--93",
isbn="978-3-319-70609-2",
doi="10.1007/978-3-319-70609-2_8",
url="https://doi.org/10.1007/978-3-319-70609-2_8"
}

@article{baksh2022said,
  title={Who said that? Impact of source expertise: A generations focused experiment on the perception of radio news sources’ gender, ethos and expertise},
  author={Baksh, Sufyan M and Fisher, Howard and Magee, Sara},
  journal={KOME: AN INTERNATIONAL JOURNAL OF PURE COMMUNICATION INQUIRY},
  volume={10},
  number={2},
  pages={76--92},
  year={2022},
  publisher={Magyar Kommunik{\'a}ci{\'o}tudom{\'a}nyi T{\'a}rsas{\'a}g}
}

@article{shane1995modern,
  title={Modern radio formats: Trends and possibilities},
  author={Shane, Ed},
  journal={J. Radio Stud.},
  volume={3},
  pages={3},
  year={1995},
  publisher={HeinOnline}
}

@article{bonini2014radio,
  title={Radio formats and social media use in Europe--28 case studies of public service practice},
  author={Bonini, Tiziano and Fesneau, Elvina and Perez, J and Luthje, Corinna and Jedrzejewski, Stanislaw and Pedroia, Albino and Rohn, Ulrike and Sellas, Toni and Starkey, Guy and Stiernstedt, Fredrik},
  journal={Radio Journal: International Studies in Broadcast \& Audio Media},
  volume={12},
  number={1-2},
  pages={89--107},
  year={2014},
  publisher={Intellect}
}

@book{warren2004radio,
  title={Radio},
  author={Warren, Steve},
  year={2004},
  publisher={Taylor \& Francis}
}

@misc{radio_2019_nielsen,
	title = {Tops of 2019: Radio},
	howpublished = {\url{https://www.nielsen.com/us/en/insights/article/2019/tops-of-2019-radio/}},    
	author = {{Nielsen}},
	note = {[Online; accessed 19-January-2022]}   
}

@misc{radioformats_glenn,
	title = {What Are Radio Formats?},
	howpublished = {\url{https://www.thebalancecareers.com/what-are-radio-formats-and-why-do-they-matter-2315430}},    
	author = {{Glenn Halbrooks}},
	note = {[Online; accessed 19-January-2022]}   
}

@misc{nielsen_rating,
	title = {NIELSEN TOPLINE RATINGS FOR SUBSCRIBING RADIO STATIONS},
	howpublished = {\url{https://tlr.nielsen.com/tlr/public/market.do?method=loadAllMarket}},    
	author = {{The Nielsen Company}},
	note = {[Online; accessed 07-August-2023]}   
}

@book{ala2005saleable,
  title={Saleable compromises. Quality cultures in Finnish and US commercial radio},
  author={Ala-Fossi, Marko},
  year={2005},
  publisher={Tampere University Press}
}

@article{sweeting2006coordination,
  title={Coordination, differentiation, and the timing of radio commercials},
  author={Sweeting, Andrew},
  journal={Journal of Economics \& Management Strategy},
  volume={15},
  number={4},
  pages={909--942},
  year={2006},
  publisher={Wiley Online Library}
}

@article{generali2011happens,
  title={What Happens When the Spots Come On? 2011 Edition},
  author={Generali, P and Kurtzman, W and Rose, B},
  journal={Coleman Insights},
  year={2011}
}

@article{crider2020tomatoes,
  title={Of “Tomatoes” and Men: A Continuing Analysis of Gender in Music Radio Formats},
  author={Crider, David},
  journal={Journal of Radio \& Audio Media},
  volume={27},
  number={1},
  pages={134--150},
  year={2020},
  publisher={Taylor \& Francis}
}

@article{AudioandTextSentimentAnalysisofRadioBroadcasts,
author = {{Dhariwal, Naman and Akunuri, Sri and Shivama, and Kather, Sharmila}},
year = {2023},
month = {01},
pages = {1-1},
title = {Audio and Text Sentiment Analysis of Radio Broadcasts},
volume = {PP},
journal = {IEEE Access},
doi = {10.1109/ACCESS.2023.3331226}
}

@Article{knowledge2030020,
AUTHOR = {Kotsakis, Rigas and Dimoulas, Charalampos},
TITLE = {Extending Radio Broadcasting Semantics through Adaptive Audio Segmentation Automations},
JOURNAL = {Knowledge},
VOLUME = {2},
YEAR = {2022},
NUMBER = {3},
PAGES = {347--364},
URL = {https://www.mdpi.com/2673-9585/2/3/20},
ISSN = {2673-9585},
DOI = {10.3390/knowledge2030020}
}

@misc{álvarez2024radiaradioadvertisement,
      title={RADIA -- Radio Advertisement Detection with Intelligent Analytics}, 
      author={Jorge Álvarez and Juan Carlos Armenteros and Camilo Torrón and Miguel Ortega-Martín and Alfonso Ardoiz and Óscar García and Ignacio Arranz and Íñigo Galdeano and Ignacio Garrido and Adrián Alonso and Fernando Bayón and Oleg Vorontsov},
      year={2024},
      eprint={2403.03538},
      archivePrefix={arXiv},
      primaryClass={cs.SD},
      url={https://arxiv.org/abs/2403.03538}, 
}

@inproceedings{bredin2020pyannote,
  title={Pyannote.audio: neural building blocks for speaker diarization},
  author={Bredin, Herv{\'e} and Yin, Ruiqing and Coria, Juan Manuel and Gelly, Gregory and Korshunov, Pavel and Lavechin, Marvin and Fustes, Diego and Titeux, Hadrien and Bouaziz, Wassim and Gill, Marie-Philippe},
  booktitle={ICASSP 2020-2020 IEEE International Conference on Acoustics, Speech and Signal Processing (ICASSP)},
  pages={7124--7128},
  year={2020},
  organization={IEEE}
}

@misc{speechbrain,
  title={{SpeechBrain}: A General-Purpose Speech Toolkit},
  author={Mirco Ravanelli and Titouan Parcollet and Peter Plantinga and Aku Rouhe and Samuele Cornell and Loren Lugosch and Cem Subakan and Nauman Dawalatabad and Abdelwahab Heba and Jianyuan Zhong and Ju-Chieh Chou and Sung-Lin Yeh and Szu-Wei Fu and Chien-Feng Liao and Elena Rastorgueva and François Grondin and William Aris and Hwidong Na and Yan Gao and Renato De Mori and Yoshua Bengio},
  year={2021},
  eprint={2106.04624},
  archivePrefix={arXiv},
  primaryClass={eess.AS},
  note={arXiv:2106.04624}
}

@software{brian_mcfee_2022_6097378,
  author       = {Brian McFee and
                  Alexandros Metsai and
                  Matt McVicar and
                  Stefan Balke and
                  Carl Thomé and
                  Colin Raffel and
                  Frank Zalkow and
                  Ayoub Malek and
                  Dana and
                  Kyungyun Lee and
                  Oriol Nieto and
                  Dan Ellis and
                  Jack Mason and
                  Eric Battenberg and
                  Scott Seyfarth and
                  Ryuichi Yamamoto and
                  viktorandreevichmorozov and
                  Keunwoo Choi and
                  Josh Moore and
                  Rachel Bittner and
                  Shunsuke Hidaka and
                  Ziyao Wei and
                  nullmightybofo and
                  Adam Weiss and
                  Darío Hereñú and
                  Fabian-Robert Stöter and
                  Pius Friesch and
                  Matt Vollrath and
                  Taewoon Kim and
                  Thassilo},
  title        = {librosa/librosa: 0.9.1},
  month        = {feb},
  year         = {2022},
  publisher    = {Zenodo},
  version      = {0.9.1},
  doi          = {10.5281/zenodo.6097378},
  url          = {https://doi.org/10.5281/zenodo.6097378}
}

@misc{Plakal_Ellis, title={YAMNet}, url={https://github.com/tensorflow/models/tree/master/research/audioset/yamnet}, journal={GitHub}, author={Plakal, Manoj and Ellis, Dan}}

@misc{radio_garden,
           month = {December},
           title = {radio.garden},
          author = {Caroline Mitchell and Lewis Peter and F{\"o}llmer Golo and Jauert Per and Kreutzfeldt Jacob and Badenoch Alexander and de Leeuw Sonja},
            year = {2016},
             url = {http://sure.sunderland.ac.uk/id/eprint/9477/},
        
}

@article{OBrien18082019,
author = {Anne O’Brien},
title = {Women in community radio: a framework of gendered participation},
journal = {Feminist Media Studies},
volume = {19},
number = {6},
pages = {787--802},
year = {2019},
publisher = {Routledge},
doi = {10.1080/14680777.2018.1508051},
URL = {https://doi.org/10.1080/14680777.2018.1508051},
eprint = {https://doi.org/10.1080/14680777.2018.1508051}
}

@article{10_1093_cdj_bsz030,
    author = {Rimmer, Annette},
    title = {Breaking the silence: community radio, women, and empowerment},
    journal = {Community Development Journal},
    volume = {56},
    number = {2},
    pages = {338-355},
    year = {2020},
    month = {01},
    issn = {0010-3802},
    doi = {10.1093/cdj/bsz030},
    url = {https://doi.org/10.1093/cdj/bsz030},
    eprint = {https://academic.oup.com/cdj/article-pdf/56/2/338/37196043/bsz030.pdf},
}

@article{Kagan2020,
  author    = {Kagan, Dima and Chesney, Thomas and Fire, Michael},
  title     = {Using Data Science to Understand the Film Industry’s Gender Gap},
  journal   = {Palgrave Communications},
  year      = {2020},
  volume    = {6},
  number    = {1},
  pages     = {92},
  doi       = {10.1057/s41599-020-0436-1},
  url       = {https://doi.org/10.1057/s41599-020-0436-1},
  issn      = {2055-1045}
}

@article{asr2021gender,
  title={The gender gap tracker: Using natural language processing to measure gender bias in media},
  author={Asr, Fatemeh Torabi and Mazraeh, Mohammad and Lopes, Alexandre and Gautam, Vagrant and Gonzales, Junette and Rao, Prashanth and Taboada, Maite},
  journal={PloS one},
  volume={16},
  number={1},
  pages={e0245533},
  year={2021},
  publisher={Public Library of Science San Francisco, CA USA}
}

@article{Williamson03072021,
author = {Patricia A. Williamson and Ethan A. Kolek},
title = {The Underrepresentation of Women on Commercial FM-Radio Stations in the Top 20 Markets},
journal = {Journal of Radio \& Audio Media},
volume = {28},
number = {2},
pages = {307--326},
year = {2021},
publisher = {Routledge},
doi = {10.1080/19376529.2020.1751632},
URL = {https://doi.org/10.1080/19376529.2020.1751632},
eprint = {https://doi.org/10.1080/19376529.2020.1751632}
}

@article{pelloin2024automatic,
  title={Automatic classification of news subjects in broadcast news: Application to a gender bias representation analysis},
  author={Pelloin, Valentin and Dodson, Lena and Chapuis, {\'E}mile and Herv{\'e}, Nicolas and Doukhan, David},
  journal={arXiv preprint arXiv:2407.14180},
  year={2024}
}

@article{laor2022radio,
  title={Radio on demand: New habits of consuming radio content},
  author={Laor, Tal},
  journal={Global media and communication},
  volume={18},
  number={1},
  pages={25--48},
  year={2022},
  publisher={SAGE Publications Sage UK: London, England}
}

@article{Sujoko31122023,
author = {Anang Sujoko, Dyan Rahmiati and Fathur Rahman},
title = {The role of radio as the public sphere for public political education in the digital era: Challenges and pitfalls},
journal = {Cogent Social Sciences},
volume = {9},
number = {1},
pages = {2239627},
year = {2023},
publisher = {Cogent OA},
doi = {10.1080/23311886.2023.2239627},
URL = { https://doi.org/10.1080/23311886.2023.2239627},
eprint = {https://doi.org/10.1080/23311886.2023.2239627}
}

@article{purwins2019deep,
  title={Deep learning for audio signal processing},
  author={Purwins, Hendrik and Li, Bo and Virtanen, Tuomas and Schl{\"u}ter, Jan and Chang, Shuo-Yiin and Sainath, Tara},
  journal={IEEE Journal of Selected Topics in Signal Processing},
  volume={13},
  number={2},
  pages={206--219},
  year={2019},
  publisher={IEEE}
}

@article{JAHANGIR2021114591,
title = {Speaker identification through artificial intelligence techniques: A comprehensive review and research challenges},
journal = {Expert Systems with Applications},
volume = {171},
pages = {114591},
year = {2021},
issn = {0957-4174},
doi = {https://doi.org/10.1016/j.eswa.2021.114591},
url = {https://www.sciencedirect.com/science/article/pii/S0957417421000324},
author = {Rashid Jahangir and Ying Wah Teh and Henry Friday Nweke and Ghulam Mujtaba and Mohammed Ali Al-Garadi and Ihsan Ali}
}

@article{SHAHFAHAD2021102951,
title = {A survey of speech emotion recognition in natural environment},
journal = {Digital Signal Processing},
volume = {110},
pages = {102951},
year = {2021},
issn = {1051-2004},
doi = {https://doi.org/10.1016/j.dsp.2020.102951},
url = {https://www.sciencedirect.com/science/article/pii/S1051200420302967},
author = {Md. {Shah Fahad} and Ashish Ranjan and Jainath Yadav and Akshay Deepak},
}

@article{forbes2024radio,
    title={Americans Listen To Far More Radio Than Podcasts—Even Young People, New Data Shows},
    author={Mary Whitfill Roeloffs, Forbes Staff},
    journal={Forbes},
    year={2024},
    note={Online article, accessed January 2025},
    url={https://www.forbes.com/sites/maryroeloffs/2024/04/30/americans-listen-to-far-more-radio-than-podcasts-even-young-people-new-data-shows//}
}

@article{sheikh2022introducing,
  title={Introducing ECAPA-TDNN and Wav2Vec2. 0 embeddings to stuttering detection},
  author={Sheikh, Shakeel Ahmad and Sahidullah, Md and Hirsch, Fabrice and Ouni, Slim},
  journal={arXiv preprint arXiv:2204.01564},
  year={2022}
}

@article{reid2022development,
  title={Development of a machine-learning based voice disorder screening tool},
  author={Reid, Jonathan and Parmar, Preet and Lund, Tyler and Aalto, Daniel K and Jeffery, Caroline C},
  journal={American Journal of Otolaryngology},
  volume={43},
  number={2},
  pages={103327},
  year={2022},
  publisher={Elsevier}
}

@article{kumar2016efficient,
  title={Efficient feature extraction for fear state analysis from human voice},
  author={Kumar, Palo Hemanta and Mohanty, Mihir N},
  journal={Indian Journal of Science \& Technology},
  volume={9},
  number={38},
  pages={1--11},
  year={2016}
}

@Article{app12010327,
AUTHOR = {Luna-Jiménez, Cristina and Kleinlein, Ricardo and Griol, David and Callejas, Zoraida and Montero, Juan M. and Fernández-Martínez, Fernando},
TITLE = {A Proposal for Multimodal Emotion Recognition Using Aural Transformers and Action Units on RAVDESS Dataset},
JOURNAL = {Applied Sciences},
VOLUME = {12},
YEAR = {2022},
NUMBER = {1},
ARTICLE-NUMBER = {327},
URL = {https://www.mdpi.com/2076-3417/12/1/327},
ISSN = {2076-3417},
DOI = {10.3390/app12010327}
}

@INPROCEEDINGS{5670700,
  author={Nguyen, Phuoc and Tran, Dat and Xu Huang and Sharma, Dharmendra},
  booktitle={International Conference on Communications and Electronics 2010}, 
  title={Automatic classification of speaker characteristics}, 
  year={2010},
  volume={},
  number={},
  pages={147-152},
  doi={10.1109/ICCE.2010.5670700}}

@inproceedings{muller2006automatic,
  title={Automatic recognition of speakers' age and gender on the basis of empirical studies},
  author={M{\"u}ller, Christian},
  booktitle={Ninth International Conference on Spoken Language Processing},
  year={2006}
}

@inproceedings{shafran2003voice,
  title={Voice signatures},
  author={Shafran, Izhak and Riley, Michael and Mohri, Mehryar},
  booktitle={2003 IEEE workshop on automatic speech recognition and understanding (IEEE Cat. No. 03EX721)},
  pages={31--36},
  year={2003},
  organization={IEEE}
}

@inproceedings{meignier2010lium,
  title={LIUM SpkDiarization: an open source toolkit for diarization},
  author={Meignier, Sylvain and Merlin, Teva},
  booktitle={CMU SPUD Workshop},
  year={2010}
}

@inproceedings{povey2011kaldi,
  title={The Kaldi speech recognition toolkit},
  author={Povey, Daniel and Ghoshal, Arnab and Boulianne, Gilles and Burget, Lukas and Glembek, Ondrej and Goel, Nagendra and Hannemann, Mirko and Motlicek, Petr and Qian, Yanmin and Schwarz, Petr and others},
  booktitle={IEEE 2011 workshop on automatic speech recognition and understanding},
  year={2011},
  organization={IEEE Signal Processing Society}
}

@misc{radford2022whisper,
  doi = {10.48550/ARXIV.2212.04356},
  url = {https://arxiv.org/abs/2212.04356},
  author = {Radford, Alec and Kim, Jong Wook and Xu, Tao and Brockman, Greg and McLeavey, Christine and Sutskever, Ilya},
  title = {Robust Speech Recognition via Large-Scale Weak Supervision},
  publisher = {arXiv},
  year = {2022},
  copyright = {arXiv.org perpetual, non-exclusive license}
}

@inproceedings{Beeferman2019, 
   title={RadioTalk: A Large-Scale Corpus of Talk Radio Transcripts},
   url={http://dx.doi.org/10.21437/Interspeech.2019-2714},
   DOI={10.21437/interspeech.2019-2714},
   booktitle={Interspeech 2019},
   publisher={ISCA},
   author={Beeferman, Doug and Brannon, William and Roy, Deb},
   year={2019},
   month=sep, pages={564–568} }

@article{amrane2022deep,
  title={A deep hybrid model for advertisements detection in broadcast TV and radio content},
  author={Amrane, Abdesalam and Meziane, Abdelkrim and Rezgui, Abdelmounaam and Lebal, Abdelhamid},
  journal={International Journal of Computational Vision and Robotics},
  volume={12},
  number={4},
  pages={397--410},
  year={2022},
  publisher={Inderscience Publishers (IEL)}
}

@book{hutchby2005media,
  title={Media talk: Conversation analysis and the study of broadcasting: Conversation analysis and the study of broadcasting},
  author={Hutchby, Ian},
  year={2005},
  publisher={McGraw-Hill Education (UK)}
}

@book{hutchby2013confrontation,
  title={Confrontation talk: Arguments, asymmetries, and power on talk radio},
  author={Hutchby, Ian},
  year={2013},
  publisher={Routledge}
}

@article{FURNER2021115236,
title = {Knowledge discovery and visualisation framework using machine learning for music information retrieval from broadcast radio data},
journal = {Expert Systems with Applications},
volume = {182},
pages = {115236},
year = {2021},
issn = {0957-4174},
doi = {https://doi.org/10.1016/j.eswa.2021.115236},
url = {https://www.sciencedirect.com/science/article/pii/S0957417421006680},
author = {Michael Furner and Md Zahidul Islam and Chang-Tsun Li},
}

@phdthesis{brannon2020mapping,
  title={Mapping US talk radio: a textual survey at scale},
  author={Brannon, William William Walker},
  year={2020},
  school={Massachusetts Institute of Technology}
}

@article{mittal2024wavepulse,
  title={WavePulse: Real-time Content Analytics of Radio Livestreams},
  author={Mittal, Govind and Gupta, Sarthak and Wagle, Shruti and Chopra, Chirag and DeMattee, Anthony J and Memon, Nasir and Ahamad, Mustaque and Hegde, Chinmay},
  journal={arXiv preprint arXiv:2412.17998},
  year={2024}
}

@article{doukhan2024gender,
  title={Gender Representation in TV and Radio: Automatic Information Extraction methods versus Manual Analyses},
  author={Doukhan, David and Dodson, Lena and Conan, Manon and Pelloin, Valentin and Clamouse, Aur{\'e}lien and Lepape, M{\'e}lina and Van Hille, G{\'e}raldine and M{\'e}adel, C{\'e}cile and Coulomb-Gully, Marl{\`e}ne},
  journal={arXiv preprint arXiv:2406.10316},
  year={2024}
}

@misc{koushnir2025vanpyvoiceanalysisframework,
      title={VANPY: Voice Analysis Framework}, 
      author={Gregory Koushnir and Michael Fire and Galit Fuhrmann Alpert and Dima Kagan},
      year={2025},
      eprint={2502.17579},
      archivePrefix={arXiv},
      primaryClass={cs.SD},
      url={https://arxiv.org/abs/2502.17579}, 
}

@misc{Nielsen_2025_Q2,
  title={The Record: Q2 U.S. audio listening trends},
  url={https://www.nielsen.com/insights/2025/the-record-q2-audio-listening-trends-2/},
  journal={Nielsen},
  year={2025},
  month={Jul}
}

@misc{lewis2019bart,
  title        = {{BART}: Denoising Sequence-to-Sequence Pre-training for Natural Language Generation, Translation, and Comprehension},
  author       = {Lewis, Mike and Liu, Yinhan and Goyal, Naman and Ghazvininejad, Marjan and Mohamed, Abdelrahman and Levy, Omer and Stoyanov, Veselin and Zettlemoyer, Luke},
  howpublished = {\url{https://arxiv.org/abs/1910.13461}},
  note         = {arXiv:1910.13461},
  year         = {2019}
}

@misc{anthropic2023claude,
  title        = {Claude Sonnet 3.5 (Large Language Model)},
  author       = {{Anthropic}},
  howpublished = {\url{https://claude.ai/}},
  note         = {Accessed: 2025-02-15},
  year         = {2023}
}

@report{GMMP2025Highlights,
  author = {{Global Media Monitoring Project}},
  title = {GMMP 2025 Highlights of Findings: Progress on a Plateau},
  institution = {World Association for Christian Communication},
  year = {2025},
  month = sep,
  url = {https://www.unwomen.org/sites/default/files/2025-09/gmmp2025-highlights-of-findings_03092025.pdf}
}

\end{document}